\documentstyle[eqsecnum,aps,prd]{revtex}
\begin{document}
\draft
\preprint{DAMTP-R97/12}
\title{Can COBE see the shape of the universe?}
\author{ Neil J. Cornish$^{\dagger}$, David Spergel$^{\star}$ and
Glenn Starkman$^{\ddag}$}
\address{$^{\dagger}$Relativity Group, DAMTP, Cambridge University,
Silver Street, Cambridge CB3 9EW, England}
\address{$^{\star}$Department of Astrophysical Sciences, Princeton
University, Princeton, New Jersey 08544}
\address{$^{\ddag}$Department of Physics, Case Western Reserve
University, Cleveland, Ohio 44106-7079}
\twocolumn[
\maketitle
\widetext
\begin{abstract}
In recent years, the large angle COBE--DMR data have been used to place
constraints on the size and shape of certain topologically compact
models of the universe. Here we show that this approach does not work for
generic compact models. In particular, we show that compact hyperbolic
models do not suffer the same loss of large angle power seen in flat
or spherical models. This follows from applying a topological theorem to
show that generic hyperbolic three manifolds support long
wavelength fluctuations, and by taking into account the dominant role
played by the integrated Sachs-Wolfe effect in a hyperbolic universe.
\end{abstract}
]

\begin{picture}(0,0)
\put(400,230){{\large DAMTP-R97/12}}
\end{picture} 
\vspace*{-0.2 in}

\narrowtext


In 1966, Marc Kac\cite{drum} posed the question ``Can one
hear the shape of a drum?''. In recent years a similar question has
been asked in cosmology; ``Can one see the shape of the
universe?''\cite{ll}. More formally, the question can be phrased: can
we discern the global topology of the universe by
studying fluctuations in the cosmic microwave background radiation (CMB)?

With the launch of new satellites next century, and with a
careful search for matched microwave temperatures
around pairs of circles on the last scattering surface
(``topological lensing'')\cite{css1,css2}, we
should be able to answer this question in the affirmative.
In the interim, we can ask how much can be done with the 4-year
COsmic Background Explorer (COBE) Differential Microwave
Radiometer (DMR) data\cite{ben}. Poor angular resolution and low signal to
noise make the COBE data unsuitable for direct lensing studies, but
several groups\cite{igor,flat1,sss,flat2,janna} have used
the data to put constraints
on a variety of toroidal models. Here we consider how their results
might be generalised to encompass a wider class of small universe
models\footnote{A small universe is defined to be one that is
multiply-connected on scales smaller than the particle horizon.}. In
particular we will be interested in hyperbolic models since
observations suggest we live in a negatively curved universe.
Moreover, the topology scale and the curvature scale are
intimately related in hyperbolic models, whereas in a flat universe
there is no scale at which one would expect to observe the topology.
By applying a number of results pertaining to the topology of
three manifolds, and by taking into account the integrated
Sachs-Wolfe effect, we argue that {\em generic} small universe models
cannot be constrained by COBE data. Naturally, some specific models
can be constrained by COBE data, but we
argue these are the exception rather than the rule.

Small universes enjoy the same local geometry and dynamics as the
usual simply connected Friedmann-Robertson-Walker (FRW)
models\footnote{By simply connected we mean the fundamental group
$\pi_1(\Sigma)$ is trivial. Since
$\pi_1(S^3)=\pi_1(E^3)=\pi_1(H^3)={\bf I}$,
the usual FRW models are all simply connected. A multiply-connected
model has a non-trivial fundamental group.}, but display different
global characteristics. In particular, small universes have a discrete
spectrum of eigenmodes and are globally anisotropic and inhomogeneous.
In models with locally spherical or euclidean geometry the eigenvalue
spectrum is raised above that of the simply connected models and
there is a corresponding long wavelength cut-off. For example, the
eigenvalues of the Laplacian on flat euclidean space take all values
in the range $k\in [0,\infty)$, corresponding to wavelengths
$\lambda=2\pi/k\in (\infty,0]$. However, if we compactify this space
by making the identifications $(x,y,z)=(x+n_x L, y+n_y L, z +n_z L)$
where the $n_i$ are integers, the eigenvalue spectrum becomes
discrete, $k_n=2\pi/L\, (n_x^2+n_y^2+n_z^2)^{1/2}$, and the bottom of the
spectrum is raised from $k=0$ to $k=2\pi/L$. There is then a
corresponding long wavelength cut-off $\lambda_{\rm max}=L$.

Assuming that
temperature fluctuations in the CMB are caused by
density fluctuations on the last scattering surface, this long
wavelength cut-off is translated into a suppression of large
angle power\cite{igor,flat1}. The cut-off in long wavelength power that
occurs in euclidean space was first used by Sokolov\cite{igor}
to show that a flat universe with toroidal spatial sections could not
be much smaller than the horizon size. He argued that the topology
scale had to be large enough to allow the wavelengths needed to produce
the quadrapole anisotropy measured by COBE. A number of
groups\cite{flat1,sss,flat2} have since improved on Sokolov's
bound and extended his analysis to include other flat topologies.
Recently, Levin {\it et al.}\cite{janna} have generalised these bounds
to include a non-compact, infinite volume hyperbolic topology
describing a toroidal horn.

There has been a tendency to draw general conclusions from these
few examples. Indeed, the small universe idea was declared dead in
Ref.~\cite{sss}. While it is fair to say that
positively curved small universes, and the
simplest toroidal flat universes
with topology scale much less than the horizon scale
are effectively ruled out\cite{flat2}, 
we show that the same cannot be said about negatively-curved models.
Lessons learned in flat space do not always apply in hyperbolic space. For
example, the eigenvalue spectrum is typically {\em lowered}, rather than
raised by making hyperbolic space compact, {\it ie.} $k^2$ can be
less than zero. Consequently, there need not be
a long wavelength cut-off. Even if there were, 
and even assuming a simple initial power spectrum at large
wavelengths, the existence of large angle power as measured by COBE--DMR 
still could not directly be used to constrain compact
hyperbolic models since the large angle power in a negatively curved universe
does not come from the last scattering surface. The bulk of the large
angle power is due to the decay of curvature perturbations along the
line of sight\cite{lyth,l-k,sk}. If the universe is hyperbolic, COBE
has been detecting fluctuations produced at moderate redshifts $z < 5 $,
rather than $z\sim 1200$. Consequently, the large angle power is
produced by fluctuations occuring on small comoving length scales that
only appear large due to their relatively close proximity.

In section I we briefly discuss
constraints on small universe models based on searches for ghost
images. In section II we emphasise
the importance of the ISW effect for calculating
CMB fluctuations on large angular scales. In section III we describe
the eigenmodes of infinite hyperbolic space. Section IV contains an
introduction to compact hyperbolic space and its underlying
mathematical structure. In section V we obtain
lower limits on the wavelength of the longest wavelength mode,
showing that modes with wavelengths longer than the
curvature scale usually exist, 
though in what multiplicity we cannot say. In section VI we digress
to consider a particular class of topologies closely related
to the horn topology studied by Levin {\it et al.}\cite{janna}.
In section VII we speculate about the form
of the primordial power spectrum in compact models, and describe how
the mixing property of compact hyperbolic space tends to spread power
across a wide range of angular scales.
Our conclusions can be found in section
VIII. A glossary of mathematical terms is included in the appendix.
References to words appearing in the glossary are indicated in the
text by roman superscripts, {\it eg.} betti number$^a$.

Throughout the paper we will be assuming that a cosmological
constant does not provide a significant contribution to the density
of the universe.

\includegraphics{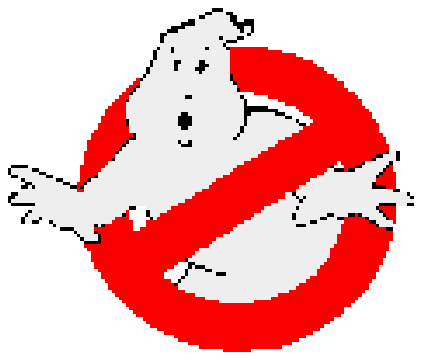}
\includegraphics{ghost.ps}

\section{Ghost Hunting}

The most obvious observational signature of a multiply connected
universe would be repeated or ``ghost'' images of familiar objects
such as galaxies or rich clusters\cite{ghost}. However, searches for
ghost images are
hampered by evolution of the objects; our ability to recognise
objects when viewed from different directions; and the difficulty
in determining the distances to objects.

Despite these problems, the consensus seems to be that there is no
evidence for ghost images out to redshifts of $z\sim 0.4$ -- the
current depth of wide-field redshift surveys. It is interesting to
note that this lack of ghost images is exactly what one expects
for typical small compact hyperbolic models. According to
Thurston\cite{bill}, the expectation value for the length of the
shortest closed geodesics in a typical small hyperbolic universe is
roughly $(0.5 \rightarrow 1.0)R_{0}$, where
$R_{0}=H_0^{-1}/\sqrt{1-\Omega_0}$  is the comoving
curvature radius. Here $H_0$ is the Hubble constant and $\Omega_0$ is the
matter density in units of the critical density. The first copies of
the Milky Way galaxy or Coma cluster would not be seen
before a conformal lookback time of $\eta\simeq 0.5 \rightarrow 1.0$.
Converting this to redshift space via the relation
\begin{equation}
1+z=\frac{2(\Omega^{-1}_0-1)}{\cosh (\eta _0-\eta )-1} \, ,
\end{equation}
where $\eta_0={\rm arccosh}(2/\Omega_0 -1)$ is the present conformal
time, we find that the first ghost images will be at a redshift of
$z\simeq 0.9 \rightarrow 2.9$ in a universe with $\Omega_0=0.3$. If
the universe has $\Omega_0$ closer to unity, the first ghost images
will be even more distant.

These numbers suggest that direct searches for ghost images of
astrophysical objects will be unable to tell if we live in a compact
hyperbolic universe. A more promising approach is to look for
topological lensing of the last scattering surface by studying
fluctuations in the cosmic microwave background radiation\cite{css1,css2}.

\section{Microwave Background Fluctuations}

Conventional lore holds that the finite size of a small universe
will lead to a long wavelength cut-off in the spectrum of primordial
fluctuations. In the sections to follow, we show that this is not guaranteed
in a hyperbolic universe. In this section we point out that even if
there were such a cut-off, it would be masked by the integrated
Sachs-Wolfe (ISW) effect on the angular scales probed by COBE.

In adiabatic models ({\it e.g.} inflation) the primordial fluctuation
spectrum determines the power spectrum on the last
scattering surface. However, the fluctuations measured by COBE do not
necessarily originate on the last scattering surface. In a
negatively curved universe, power on angular scales larger
than the curvature scale is produced at relatively low redshifts by
fluctuations occuring on scales considerably smaller than the
curvature scale. This severely limits COBE's ability to probe the
large scale topology of the universe.

In a hyperbolic universe, there are two terms
that produce the microwave background fluctuations 
on large angular scales:
\begin{equation}\label{isw}
{\Delta T (\theta,\phi) \over T}=  {\Phi({\bf x}_{sls},\eta_{sls})  \over 3}
+ 2 \int_{\eta_{sls}}^{\eta_0} \dot \Phi ({\bf x},\eta) d\eta \, ,
\end{equation}
where $\eta_{sls}$  is the conformal time 
at the surface of last scatter and $\eta_0$ is the present conformal
time.  The first term is due to variations in the gravitational
potential and photon density at the surface of last scatter.  The latter
term, which is zero to linear order in a matter dominated 
{\it flat} universe, is due to the decay of potential fluctuations
at late times, $(1 + z) < \Omega_0^{-1}$.
In a universe with $\Omega_0 = 0.3$, the latter term,
the so-called integrated Sachs-Wolfe effect, is the
{\it dominant} source of microwave background fluctuations on large
angular scales\cite{sk}. The late-time ISW effect dominates multipole
moments below $\ell_{\rm curv}=2\sqrt{1-\Omega_0}/\Omega_0$.
Neglecting this contribution will lead to a severe underestimate of
the large angle power.

Because of the late-time ISW effects, we expect significant
large angular scale fluctuations even if $\Omega_0$ is as
small as 0.1.  Kamionkowski \& Spergel \cite{sk} calculated
the CMB fluctuations in hyperbolic models with trivial topology
and a variety of primordial power spectra. 
Below the curvature scale the standard scale-invariant
Harrison-Zeldovich spectrum was used. This is probably reasonable 
at sufficiently small scales for
models with non-trivial topology, including compact models,
as small scale perturbations will be less sensitive to global properties such
as the curvature and topology. Beyond the curvature scale,
both the unknown form of the eigenmodes and the expected 
effects of transients  in the inflationary dynamics make
the situation much less clear. 
For the latter reason, Kamionkowski \& Spergel
considered a range of power spectra. Of the models they considered, the
``volume power law model'' has the least long wavelength power. In
this model, fluctuations on scales larger than the curvature scale are
exponentially suppressed. Consequently, there is essentially no
contribution to $\Delta T /T$ from the last scattering surface in the
volume power law model. Nevertheless, the second term in (\ref{isw})
produced sufficient power to fit the fluctuations observed by
COBE--DMR (see figures 6 and 9 in \cite{sk}).

The above result is not difficult to understand. As mentioned earlier, the
large angle power in a sub-critical universe is produced by
fluctuations occuring on small comoving length scales that
only appear large due to their relatively close proximity. To make
this concrete, we can consider the contribution $\alpha_\ell (k)$
to a given multipole, $\ell$, from modes with wavenumber $k$\cite{l-k}:
\begin{equation}
\alpha_\ell (k)= \Phi_k(\eta_0)\, \widetilde \alpha_\ell (k)
\end{equation}
where
\begin{eqnarray}\label{alpha}
&& {\widetilde \alpha}_\ell (k)=\left[
\frac{1}{3} F(\eta_{\rm sls})X^{\ell}_{k}(\eta_0-\eta_{sls}) \right.
\nonumber \\
&& \hspace*{0.4in} \left. +2\int_{\eta_{sls}}^{\eta_0} { d F \over d
\eta}(\tilde\eta)\, X^{\ell}_{k} (\eta_0-\tilde\eta)\,  d{\tilde\eta}
\right] \, .
\end{eqnarray}
Here $X^{\ell}_{k}$ are the radial eigenfunctions of the
Laplacian on $H^3$ (see next section), and $\Phi_k(\eta)$ describes
the curvature perturbation on scales
$2\pi /k$ at conformal time $\eta$. These are related to the curvature
perturbations today by $\Phi_k(\eta)=\Phi_k(\eta_0)F(\eta)/F(\eta_0)$
where\cite{muk}
\begin{equation}
F(\eta)= 5 \, {\sinh^2(\eta)-3 \eta \sinh\eta +4 \cosh\eta -4 \over
(\cosh \eta -1)^3} \, .
\end{equation}

\
\begin{figure}[h]
\vspace{54mm}
\includegraphics{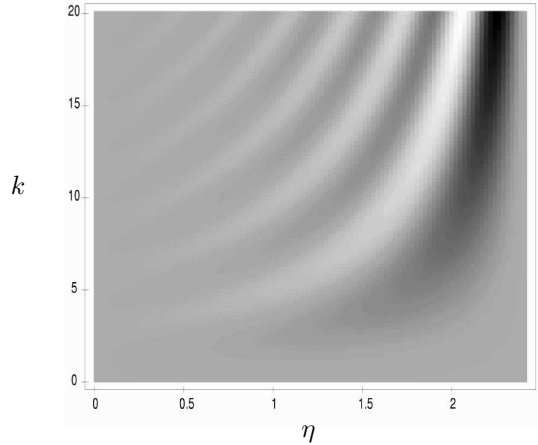}
\vspace{5mm}
\caption{A density plot showing the quadrapole integrand,
$F'(\eta)X^{2}_{k}(\eta_0-\eta)$, as a function of
wavenumber $k$ and time since last scatter $\eta$. The regions of
highest contrast are where the dominant contribution occurs.}
\end{figure} 

\vspace*{-2mm}
\begin{picture}(0,0)
\put(0,160){$k$}
\put(110,68){$\eta$}
\end{picture}

In a flat universe $dF/d\eta=0$ to leading order, and only the first
term contributes. Moreover, it is easy to show that
$\tilde\alpha_\ell (k)$ is strongly peaked at $k_{sls} \sim \ell+1$ in a flat
universe if we chose our unit of length to be the radius of the
surface of last scatter. The same is true in an open universe for
multipoles with $\ell \gg \ell_{{\rm curv}}$, but for low multipoles
the second term in (\ref{alpha}) dominates. For example, Fig.~1 shows
which modes $k$ contribute most to the quadrapole integrand,
$F'(\eta)X^{2}_{k}(\eta_0-\eta)$, in a universe
with $\Omega_0=0.3$. Notice that the dominant contribution
comes from late times, $\eta > 1.5$, $z<2$ and large wavenumber.

\begin{figure}[h]
\vspace{53mm}
\includegraphics{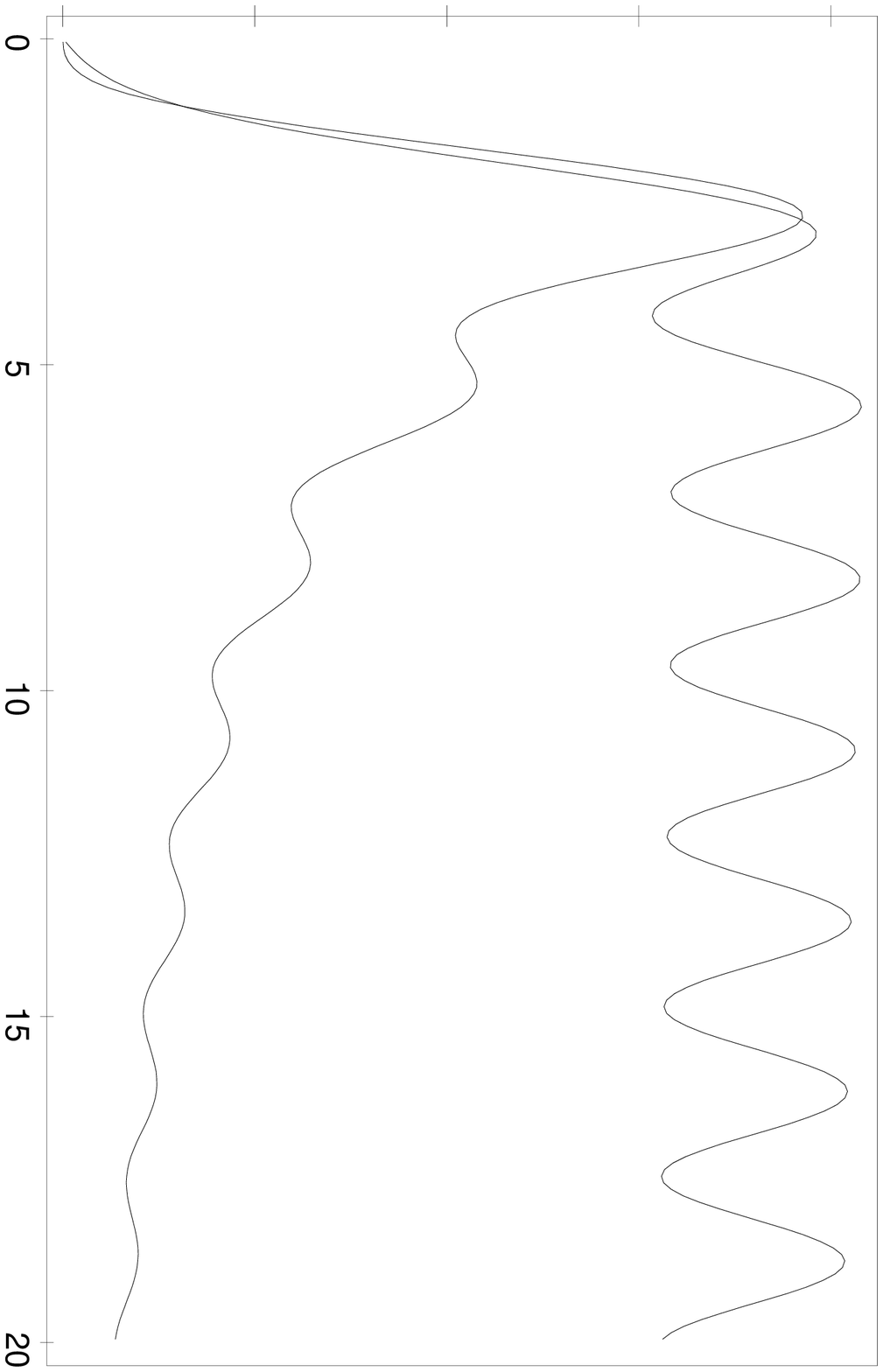}
\vspace{8mm}
\caption{The functions $|\widetilde \alpha_2 (k)|$ and $|\alpha_2
(k)|$ in a universe with $\Omega_0=0.3$. The scaling is arbitrary.}
\end{figure} 

\vspace*{-2mm}
\begin{picture}(0,0)
\put(150,160){$|\widetilde \alpha_2 (k)|$}
\put(150,90){$|\alpha_2 (k)|$}
\put(110,45){$k$}
\end{picture}

We highlight the wavenumber dependence in Fig.~2 by plotting
$|\widetilde \alpha_2 (k)|$ and $|\alpha_2 (k)|$. The latter is shown for
a volume power law scaling with spectral index $n=1$. We see that
the dominant contribution to the quadrapole comes from modes
with wavenumbers $k \sim 3 \rightarrow 10$ in curvature units or
$k_{sls} \sim 7 \rightarrow 24$ in units of the radius of the
surface of last scatter. This should be contrasted with the flat space
case where modes with $k_{sls}\sim 2 \rightarrow 4$
provide the dominate contribution to the quadrapole.

The important lesson in all this for  compact hyperbolic models is that 
the form of the power spectrum at very small wavenumber (long wavelength)
is largely irrelevant. It is power on scales smaller than the
curvature scale that
contributes most to the low multipoles. Since the topology scale is typically
comparable to or larger than the curvature scale, the ISW
effect should ensure there is no significant suppression of the large
angle temperature fluctuations in generic compact models.

In contrast to the amplitude fluctuations, microwave background
polarisation fluctuations exclusively probe
the surface of last scatter.  These fluctuations arise
due to electron scattering and depend on gradients in the
velocity field\cite{zalhar,zalsp}
\begin{eqnarray}
(Q+iU)(\hat {{\bf n}})=0.17\Delta \tau _{*}{\bf m}^i{\bf m}^j\partial
_iv_j|_{\tau _{*}}  
\end{eqnarray}
where $\Delta \tau _{*}$ is the width of the last scattering surface and is
giving a measure of the distance photons can travel between their last two
scatterings.  Here, $Q$ and $U$ are the Stokes parameters, ${\bf n}$
is the direction of photon propagation, and ${\bf m} = {\hat {\bf e_1}}
+ i {\hat {\bf e_2}}$, where ${\hat {\bf e_1}}$ and ${\hat {\bf e_2}}$
form a basis orthogonal to ${\bf n}$.  Note that polarisation
fluctuations are produced only due to scattering and are not
sensitive to the ISW effects.  Thus, polarisation fluctuations
are our best chance to directly probe supercurvature modes.

\section{Vibrations in a hyperbolic cavity}
When attempting to calculate perturbation spectra in compact
hyperbolic space one is immediately confronted by the highly
non-trivial task of finding the eigenmodes.
In principle the eigenmodes of a compact space can be obtained from
the eigenmodes of the simply connected covering space using the method
of images\cite{gutz1}. In practice the sums involved are
highly divergent and can only be tamed by sophisticated resummation
methods\cite{berry1,agam}. Before confronting this challenging
problem we need to know the eigenmodes of the covering space.
The covering space has the metric
\begin{eqnarray}\label{cov_met}
&&ds^2=dt^2-R^2(t)d\sigma^2 \, , \nonumber \\
&&\hspace{0.2in}= R^2(\eta) (d\eta^2 - d \sigma^2) \, ,
\end{eqnarray}
where the $d\sigma^2$ is the metric on hyperbolic three-space,
\begin{equation}\label{three}
d\sigma^2=d\chi^2+\sinh^2\chi\left(d\theta^2
+\sin^2\theta d\phi^2\right) \, .
\end{equation}
The Ricci curvature of this metric is $-1$, corresponding to a
curvature scale of unity.
Perturbations in such a spacetime can be expanded in terms of
spherically symmetric solutions of the Helmholtz equation
$(\Delta+q^2)Q=0$,
where the $\Delta$ is the Laplace operator on $H^3$,
\begin{eqnarray}
&&\Delta Q = {1 \over \sinh^2\chi}\left[ {\partial \over \partial \chi}
\left( \sinh^2\chi {\partial Q \over \partial \chi} \right) + \right.
\nonumber \\
&&\quad 
\left. {1 \over \sin^2\theta}{\partial \over \partial \theta}\left(
\sin\theta {\partial Q \over \partial \theta}\right)
+{1 \over \sin^2\theta}{\partial^2 Q \over \partial \theta^2}\right]
\, .
\end{eqnarray}
The eigenfunctions are given by\cite{absh}
\begin{equation}
Q^{q\ell m}(\chi,\theta,\phi) = X^{\ell}_{q}(\chi) 
Y^{m}_{\ell}(\theta,\phi) \, ,
\end{equation}
where the $Y^m_\ell$'s are spherical harmonics and the radial eigenfunctions
are given by
\begin{equation}
X^{\ell}_{q}(\chi)={ (-1)^{\ell+1} \sinh^\ell\chi\over
 \left(\prod_{n=0}^{l}(n^2+k^2)\right)^{1/2}} 
  \, {d^{\ell+1} \cos(k \chi) \over
d(\cosh\chi)^{\ell+1}} \, .
\end{equation}
The wavenumber, $k=2\pi /\lambda$, is related to the eigenvalues of
the Laplacian by
\begin{equation}
k^2=q^2-1 \, .
\end{equation}
In the literature there is considerable confusion surrounding this
shift between eigenvalue and wavenumber in hyperbolic space. Some
authors claim that $q$ is the wavenumber, but this is not
true. Indeed, it is simple to prove that for $\chi > 2\pi /k$, the
radial eigenfunctions take the form
\begin{equation}
X^{\ell}_{q}(\chi) \sim { \cos(k\chi+\phi_{k\ell}) \over \sinh\chi} \, ,
\end{equation}
where $\phi_{k\ell}$ is a $k, \ell$ dependent phase. The $1/\sinh\chi$
factor follows from flux conservation in a space where the surface
area of a ball grows as $4\pi \sinh^2\chi$. Clearly, $k$ is the
wavenumber and $\lambda=2\pi/k$ is the wavelength.
We will refrain from calling modes with $\lambda>1$ ``supercurvature''
to avoid confusion with other papers in the literature where
``supercurvature'' is used to describe modes with $q^2 < 1$.
Perhaps the confusion surrounding wavenumbers in open models
comes from considering the wave equation for massless scalar fields:
\begin{equation}
\left( { \partial^2 \over \partial \eta^2} - \Delta \right) \Psi(\eta,
{\bf x}) =0 \, .
\end{equation}
For eigenmodes $\Psi(\eta,{\bf x})$ with eigenvalue $q$ and
angular frequency $\omega_q$ we have
\begin{equation}
\omega^2_q = q^2 = k^2 +1={(2\pi)^2\over \lambda^2}+1\, .
\end{equation}
Notice that the usual relationship between frequency and wavelength
is offset by one unit. If we were to neglect this offset and assert that
$\omega=2\pi/\lambda$, then we would erroneously conclude that
$q$ was the wavenumber. 

In compact hyperbolic space the eigenmodes will be
discrete and the spectrum can be {\em lowered} below $k^2=0$.
Modes with $k^2<0$ are not square integrable in infinite hyperbolic
space as they grow exponentially with $\chi$.
However, these modes are square integrable in compact hyperbolic space
and are thus quite acceptable.

The physical and comoving counterparts to the wavenumber $k$ and
wavelength $\lambda$ are scaled such that
\begin{eqnarray}
&&k_{{\rm phys}}={k \over R(t)}\, ,\quad k_{{\rm cmvg}}={k \over R_0 }=k H_0
\sqrt{1-\Omega_0} \, , \nonumber \\
&&\lambda_{{\rm phys}}=R(t)\lambda \, ,\quad \lambda_{{\rm cmvg}}=R_0
\lambda = {\lambda \over H_0 \sqrt{1-\Omega_0} } \, .
\end{eqnarray}

Fluctuations in the temperature of the cosmic microwave background
are due to variations in the gauge invariant gravitational
potential $\Phi({\bf x},\eta)$. The connection between eigenvalue
spectra and observed fluctuations in the CMB follows from the relation
\begin{equation}
\Phi({\bf x})=\sum_{q,\ell, m} c_{q\ell m}Q^{q\ell m}({\bf x}) \, .
\end{equation}
The expansion coefficients $c_{q\ell m}$ are fixed by the primordial
power spectrum. Moreover, any physical mechanism for generating that
primordial power will be influenced by the shape of the eigenmodes and
the eigenspectrum, for no matter how skilled the drummer, a snare drum
will not sound like a timpani.

\section{Compact Hyperbolic Space}

A compact hyperbolic universe has spatial sections of the form
$\Sigma= H^3/\Gamma$, where the fundamental group, $\Gamma$, 
is a discrete subgroup of
$SO(3,1)\cong PSL(2,C)$ acting freely ({\it ie.} without fixed
points) and discontinuously (since it is discrete). According to
Poincare's fundamental polyhedron theorem\cite{ponc}, $\Sigma$ can be
obtained by gluing together the faces of a polytope in hyperbolic
space. The polytope is otherwise refered to as the manifold's
fundamental cell or Dirichlet domain\footnote{A simple analogue 
in two dimensions is the torus, $E^2/\Gamma$, where $E^2$ is the plane
and $\Gamma$ is the group generated by a translations
by $L_x$ in the $x$ direction, and $L_y$ in the $y$ direction.
The fundamental cell for this torus is a rectangle with opposite
faces identified.}.

Any function defined on the compact space
$\Sigma = H^3/\Gamma$ must be invariant under the
the action of the fundamental group $\Gamma\subset SO(3,1)$. The
simplest way to enforce this condition employs the method of images:
\begin{equation}\label{images}
Q_{\Gamma}({\bf x})=\sum_{g \in \Gamma} Q(g{\bf x}) \, .
\end{equation}
The same method can be used to generate any $n$-point function
in the compact space via a sum over translated copies of the
corresponding function in the covering space. In a recent paper,
Bond {\it et al.}\cite{bond} applied the method of images to the
two-point correlation function in several compact hyperbolic
universes. They concluded that several of the smaller volume
hyperbolic models were incompatible with the COBE data. However, in
this preliminary study they did not include the ISW effect, nor did
they demonstrate that their results are independent of the infrared
regularisation scheme they used. They\cite{Souradeep}
recently reported a new analysis that includes the ISW effect
and appears to be consistent with our conclusion that
COBE is compatible with compact manifolds.

Hyperbolic 3-space can be viewed as the unit hyperboloid (mass-shell)
\begin{equation}
-x_0^2+x_1^2+x_2^2+x_3^2=-1 \, , \label{hype}
\end{equation}
embedded in 4-dimensional Minkowski space. We can relate this
representation to the induced metric on $H^3$, (\ref{three}),
by the coordinate identifications
\begin{eqnarray}\label{coord}
 x_0=\cosh\chi\, , &\quad & x_1=\sinh\chi \cos\theta\, , \nonumber \\
 x_2=\sinh\chi\sin\theta\cos\phi\, , &\quad & x_3=\sinh\chi\sin\theta\sin\phi
\, .
\end{eqnarray}
From this perspective it is
easy to understand why the isometries of $H^3$ are described by the
orientation preserving homogeneous Lorentz group in
4-dimensions, $SO(3,1)$. Given a set of
generators $\{a_1,..,a_j\}$, any element of the
fundamental group $\Gamma$ can be written as
\begin{equation}\label{gen}
g=\prod_{i}a_{m_{i}}^{j_{i}} \quad (i,j_i,m_i \in {\bf Z}) \, ,
\end{equation}
with possible repetitions of the generators. The group element $g$ is
called a word, and the length of the word is defined to be
\begin{equation}
l_g = \sum_i |j_i| \, .
\end{equation}
Not all words generated according to (\ref{gen}) will be unique since
the generators are typically subject to a set of relations, {\it e.g.}
$a_1a_2a_1^{-2}a_2=1$. The number of distinct words with lengths less
than or equal to $l$ is denoted ${\cal N}(l)$. A theorem due to
Milnor\cite{mil} tells us that ${\cal N}(l)$ grows exponentially with
$l$ if $\Gamma$ is the fundamental group of a compact hyperbolic
manifold. It is precisely this exponential growth that causes problems
with the sum over images. The rate of growth is measured by the
grammatical complexity or topological entropy of the fundamental group,
$H_T=\lim_{\, l\rightarrow \infty} l^{-1} \log[{\cal N}(l)]$.

To illustrate the preceeding discussion we use {\it
SnapPea}\cite{weeks} to study Thurston's
manifold\cite{thur1}, $\Sigma_{{\rm Th}}$
[m003(-2,3) in the {\it SnapPea} census]. The fundamental
group, $\Gamma=\pi_1(\Sigma_{{\rm Th}})$, has the
presentation\footnote{A presentation lists the group generators followed by
any words which are equivalent to the identity.}
\begin{equation}\label{fun}
\Gamma=\{ a,b \, : \, a^2ba^{-1}b^3a^{-1}b,\;
ababa^{-1}b^{-1}ab^{-1}a^{-1}b \} \, .
\end{equation}
The generators of the fundamental group describe identifications in
the faces of the fundamental cell shown in Fig.~3. The fundamental
cell is drawn using Klein's projective model for hyperbolic space. In
this projection $H^3$ is mapped into an open ball in $E^3$. Under this
mapping hyperbolic lines and planes are mapped into their euclidean
counterparts. This is why the totally geodesic faces of the
fundamental cell appear as flat planes.

\
\begin{figure}[h]
\vspace{50mm}
\includegraphics{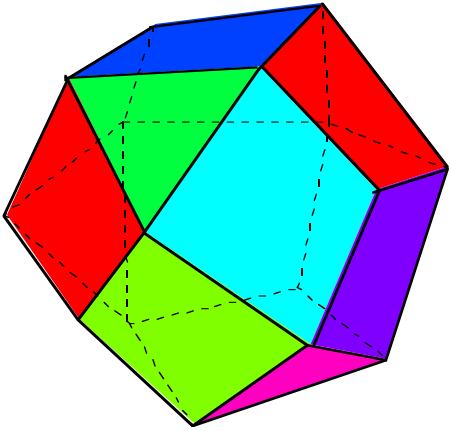}
\vspace{8mm}
\caption{The fundamental cell for Thurston's manifold.}
\end{figure} 

Thurston's manifold
has volume $0.98137$, symmetry group $G=\{u,v \, :\,  u^2, v^2,
uvuv\}={\bf Z}_2\oplus {\bf Z}_2$, first homology group ${\bf Z}_5$ and betti
numbers$^a$ $b_0=b_3=1$, $b_1=b_2=0$. The symmetry group describes the
symmetries of the manifold. The (first)
homology group\cite{nak} is the abelianised version of the fundamental
group (\ref{fun}). When abelianised, the relations obeyed by the
fundamental group collapse down:
\begin{eqnarray}
&&a^2ba^{-1}b^3a^{-1}b=1 \quad \Rightarrow \quad \tilde{b}^5=1 \,
\nonumber \\
&&ababa^{-1}b^{-1}ab^{-1}a^{-1}b=1 \quad \Rightarrow \quad
\tilde{a}=\tilde{b}^{-1} \, ,
\end{eqnarray}
leaving the homology group
\begin{equation}
H_1(\Sigma_{{\rm Th}})=\{ \tilde{b} \; : \; \tilde{b}^5=1 \} = {\bf Z}_5
\, .
\end{equation}

Choosing a coordinate system centred at a maximum of the injectivity
radius function$^b$, the generators have the $SO(3,1)$ matrix
representations
\begin{equation}
a\! = \!\left( \begin{array}{cccc}
1.4498& -0.3191 & 0.8911 & -0.4538 \\
-0.5653 & -0.5653  & -0.8911 & 0.4538 \\
0.8844 & -0.8844 & 0.8911 & -0.4538\\
0.0000 & 0.0000 & -0.4538 & -0.8911\end{array} \right)\; ,
\end{equation}
and
\begin{equation}
b\! = \!\left( \begin{array}{cccc}
2.9351& 2.4389 & -1.1390 & -0.6073 \\
0.9195 & 0.4233  & -1.1390 & -0.6073 \\
2.5987 & 2.5987  & -0.8587 & -0.5125\\
0.1255 & 0.1255 & -0.5125 & 0.8587\end{array} \right)\; .
\end{equation}
The image of any point ${\bf x} \in H^3$ can now be found by matrix
multiplication. To give an example, the origin ${\bf x}={\bf 0}$ has $\chi=0$
and corresponds to the point $[1,0,0,0]$ when embedded in four
dimensional Minkowski space. Acting on this point by
$a$ takes it to the point $[1.450,-0.565,0.884,0]$. This point has
$\chi=0.9161$, $\theta=2.1395$ and $\phi=0.0157$, and so lies a
distance $0.9161$ units away in 3-space.

Points lying on a symmetry axis of a group element will be
translated the shortest distances. Conversely, the further a point
lies from the symmetry axis of a group element, the further it is
translated by that element.
Since the fundamental group acts differently on different points,
compact hyperbolic models are not homogeneous. Nor are they isotropic
since there are prefered symmetry axes. Points on the symmetry axis
of a group element can be located by finding the eigenvectors of the
$SO(3,1)$ matrix describing the group element. The two real
eigenvectors define points on the lightcone enclosing the
hyperboloid (\ref{hype}). The line passing through
these two points defines the symmetry axis of the group element. The
intersection of this line with the hyperboloid (\ref{hype}) defines
the point in $H^3$ that is translated the shortest distance.
For example, $a$ has the two real eigenvectors
\begin{eqnarray}
&&e_{1}=[-0.7491,0.3497,-0.6563,0.0896] \, , \nonumber \\
&&e_{2}=[0.7350,0.2687,-0.6460,0.2252] \, ,
\end{eqnarray}
and the line they define in Minkowski space intersects
the hyperboloid at the point
\begin{equation}
v=[1.2428,0.2409,-0.6425,0.2716] \, .
\end{equation}
Acting on this point by $a$ leads to the image point
\begin{equation}
v_a=[1.0292,-0.1431,0.19035,0.0496] \, ,
\end{equation}
a distance\footnote{A simple way to work out the length of the
shortest geodesic connecting two points is to first perform an
$O(3,1)$ rotation of the coordinate system so that one of the points
lies at the origin of $H^3$. The proper distance between the two
points is then found by taking the arccosh of the other point's
``time'' coordinate, in accordance with (\ref{coord}).}
$0.8894$ units away in 3-space.

By acting on points lying on the symmetry axis of each group element
it is possible to compile a list of the the minimal geodesics. A
typical isometry is a corkscrew type motion, consisting
of a translation of length $L$ along a geodesic, combined with
a simultaneous rotation through an angle $\omega$ about the
same geodesic. The length and torsion can be found directly from
the eigenvalues of the group element, and are conveniently listed
by the {\em SnapPea} program\cite{weeks}.
Table~I records both the length and the torsion 
of all geodesics with $L < 2$.

Each word $g\in \Gamma$ does not necessarily produce a unique minimal
geodesic. The minimal geodesic generated by $ab^2$ has the same length
and torsion as that generated by $b$. The mapping between words and
minimal geodesics is many to one. To make the mapping one to one, the
words need to be grouped into conjugacy classes. Two words, $g$ and
$g'$ belong to the same conjugacy class if and only if they are equal
up to an isometry of $\Gamma$\footnote{{\it e.g.}, For Thurston's
manifold we have
$ab^2\sim bab \sim ababa^{-1}=b^{-1}aba^{-1}b\sim aba^{-1} \sim b$.}:
\begin{equation}
g \sim g' \quad {\rm iff}\quad g'= f^{-1} g f \, \quad (f\in \Gamma) \, .
\end{equation}
A theorem by McKean\cite{mckean} then states that there is a one to
one correspondence between conjugacy classes of the fundamental group
and the periodic geodesics. If we define $\Pi_p$ to be the set of all
geodesic loops at some point $p\in \Sigma$, endowed with the product $\gamma_a
\circ \gamma_b$ (first $\gamma_b$ then $\gamma_a$) for all
$\gamma_a,\gamma_b \in \Pi_p$, then $\Pi_p$ is isomorphic to
$\pi_1(\Sigma)=\Gamma$. This link between geodesic loops and the
fundamental group can be used to re-express the sum over images
(\ref{images}) as a sum over periodic orbits. It is this principle
that forms the basis of Gutzwiller's method\cite{gutz1} for determining the
eigenmodes on compact hyperbolic space. Indeed, many of the techniques used to
describe quantum chaos, including Gutzwiller's trace formula,
have been developed using compact hyperbolic space as a
laboratory\cite{balazs}. The exponential growth
in the number of words with symbolic lengths $l_g \leq l$
is echoed by the exponential growth in the number of closed geodesics with
physical lengths $L_\gamma \leq L$:
\begin{equation}
N(L)\sim {e^{h L} \over hL} \, ,
\end{equation}
where $h$ is the Kolmogorov-Sinai (KS) entropy of the geodesic
flow\cite{balazs}. It is interesting to note that KS-entropy scales as
$h \propto V^{-1/3}$ while the topological entropy scales as
$H_T \propto V$. This is because the KS-entropy measures the
rate of chaotic mixing, and smaller manifolds mix better, while the topological
entropy measures the complexity of the fundamental group, and larger
manifolds have more complicated topologies\cite{mf}.

We can make some general observations about the existence of long
wavelength modes on $H^3/\Gamma$ based on {\em SnapPea}'s listing of the
short minimal geodesics. Typical

\begin{table}
\caption{Minimal geodesics shorter than 2}
\begin{tabular}{lll}
Length&Torsion&Word\\
0.57808244 & 2.13243064 & $ab$ \\
0.72156837 & -1.15121299 & $b$ \\
0.889442997 & 2.94185905 & $a$ \\
0.998325189 & -2.92101779 & $ab^{-1}$ \\
1.040315125 & 0.98237189 & $aba^{-1}b$ \\
1.793800843 & -1.55687105 & $a^2b$ \\
1.822279900 & -2.41353903 & $ab^{-1}a^{-1}b$
\end{tabular}
\end{table}

\noindent  closed geodesics, such as those
listed in Table.~I for Thurston's manifold, involve a considerable
torsion. A similar twisting occurs in 5 of the 6 compact, orientable
flat three manifolds\cite{wolf}. One example is $T^3_{1,\pi}$, where 
opposite faces of a cube are identified, with one pair of faces
identified after a twist through $\pi$. If
the cube has side length $L$, then the twisted minimal geodesic has length
$L$ and torsion $\pi$. As a consequence of this torsion, the lowest
eigenmode along the twisted direction must wrap twice around
$T^3_{1,\pi}$ before closing. The
maximum allowed wavelength is thus $2L$, not $L$. We may anticipate a
similar phenomenon occuring in hyperbolic space. The shortest
geodesic listed in Table.~I has torsion $2.132431 \approx
2\pi/2.9465$. This geodesic approximately closes after 3 turns, but
may never close exactly if it is an irrational multiple of $2\pi$. Using
this Bohr-Sommerfeld style reasoning, it appears likely that compact
hyperbolic manifolds will admit very long wavelength modes.

\section{Long wavelength modes}

Here we study long wavelength modes in small hyperbolic universes.
We do this without explicitly
solving for the eigenmodes by exploiting the close connection between
eigenvalue spectra and topology. We find a number of
useful topological results pertaining to long wavelength modes.

For hyperbolic manifolds of dimension $d\geq3$ there is a
remarkable connection between geometry and topology. The rigidity
theorem of Mostow-Prasad\cite{mp} proves that any connected and orientable
manifold of dimension $d\geq 3$ supports at most one hyperbolic metric
(up to diffeomorphisms). This means that geometrical quantities such as
volume, injectivity radius$^c$, diameter$^d$, geodesic length spectra
and eigenvalue spectra are all {\em topological invariants} for
compact hyperbolic manifolds.

In this section we will put the topologists' interest in the
eigenvalue spectra to good use. Without having to solve for the eigenmodes
explicitly we can prove several results concerning the existence of
long wavelength modes in compact hyperbolic spaces. In particular,
we prove that generic compact hyperbolic spaces admit modes with
wavelengths that exceed the curvature scale. In addition, we
show that there exist finite volume, compact hyperbolic manifolds
with an arbitrarily large number of modes with arbitrarily long
wavelengths. 

To relate these results to cosmology we need to recall the
relationship between curvature, redshift, density and the radius of
the surface of last scatter (SLS) in
a hyperbolic universe. The curvature radius is fixed by the scale
factor $R(t)$ since the metric (\ref{three}) has unit curvature
radius. The radius of the last scattering surface
at redshift $z$ is given by
\begin{equation}
R_{sls}=R\, {\rm arccosh}\left(1 + {2(1-\Omega_0) \over \Omega_0(1+z)}
\right)\equiv R\, \chi_{sls} \, .
\end{equation}
The volume of space encompassed by the SLS is
\begin{equation}\label{vlss}
V_{sls} = \pi R^3 (\sinh(2\chi_{sls})-2\chi_{sls}) \, .
\end{equation}
The radius of the last scattering surface today is approximately equal to
the curvature radius if $\Omega_0=0.8$. If $\Omega_0=0.4$ we find
$R_{sls} \approx 2 R_0$; if $\Omega_0=0.1$ we find
$R_{sls} \approx 3.6 R_0$. The angle subtended by the curvature
scale on the last scattering surface is approximately
\begin{equation}\label{thetac}
\theta_{{\rm curv}}\approx 1.68 {\Omega_0 \over \sqrt{1-\Omega_{0}}}
\, .
\end{equation}
The above expression assumes that the universe has been matter
dominated since decoupling. This will be true if matter-radiation
equality was reached before decoupling so that
\begin{equation}
z_{{\rm eq}}=24000\Omega_0 h^2 > z_{sls} \quad \Rightarrow
\quad \Omega_0 h^2 > 0.052 \, .
\end{equation}
Assuming $h>0.5$, (\ref{thetac}) will be valid so long as
$\Omega_0 > 0.2$. Since, roughly speaking, the $\ell^{\rm th}$ multipole moment
measures power on angular scales\footnote{The reasoning
being that the $\ell^{\rm th}$ multipole has $2\ell$ zeros in the range
$\theta\in [-\pi,\pi]$, with approximately equal spacings of
$\Delta \theta=\pi/\ell$.} of $\pi/\ell$, modes with $\lambda>1$
probe angular scales $\ell < \ell_{{\rm curv}}$, where
\begin{equation}
\ell_{{\rm curv}} \approx { 2 \sqrt{1-\Omega_0} \over \Omega_0} \, .
\end{equation}
In a universe with $\Omega_0 =0.5$, only the $\ell=2$ quadrapole probes
modes with $\lambda>1$, while in a universe with $\Omega_0=0.3$ the
range is increased to include all multipoles below $\ell=6$. This tells
us that perturbations with wavelengths larger than the curvature scale
are responsible for the large angle power on the last scattering
surface if $\Omega_0 < 0.5$.

Using (\ref{vlss}) we can estimate the redshift when a
fundamental cell first dropped within the last scattering
surface from the relation
\begin{equation}
1+z={2 (\Omega_0^{-1}-1) \over \Omega_0(\cosh r_+ -1)}\, ,
\end{equation}
where $r_+$ is the outradius$^e$ of the manifold. Taking Thurston's
manifold\cite{thur1} (see Fig.~3) with
${\rm Vol}(\Sigma_{{\rm Th}})=0.98137$ and
$r_+= 0.748537$ as a particular example, we find that
the fundamental cell dropped inside the SLS no earlier
than $z=9.2$ if $\Omega_0=0.4$.
Today there would be approximately 86 copies of
the fundamental cell within the SLS (this is the ratio of volume of
the optically observable universe to the comoving volume of Thurston's
manifold). Since the volume of a hyperbolic manifold is a measure of
topological complexity, Thurston's manifold ranks as one of the
simplest topologies the universe can have. The only known example
that is simpler is the Weeks manifold with volume $0.9427..$. It is
thought that the Weeks manifold is {\em the} smallest hyperbolic
three manifold, though the most recent lower bound, ${\rm Vol}(\Sigma_{\rm
min}) > 0.166$\cite{gabai} still leaves some room for smaller, simpler
topologies.

Returning to our treatment of the eigenvalues, we introduce the
ordering $0= q_0 < q_1 < q_2 \dots $, where the eigenvalues are counted with
their multiplicities. The mathematical literature is littered with
dozens of upper and lower bounds for the $q_j$'s in terms of the
volume, diameter or isoperimetric constant of a
manifold. Unfortunately most of these bounds are not very sharp since
the results apply to a great variety of manifolds. Sharper bounds can
probably be found by restricting ones attention to three dimensional
manifolds with constant negative curvature. 

Most papers deal with the first eigenvalue, $q_1$, whereas we are most
interested in eigenmodes with $q^2 \in [1,1+\epsilon^2]$ where $\epsilon
\ll 1$. Eigenmodes in this interval correspond to modes with
wavelengths $\lambda \geq 2\pi/\epsilon$. Nevertheless, some of the
bounds on $q_1$ are useful to us.

Many of the bounds on $q_1$ employ Cheeger's isoperimetric
constant\cite{cheeg1}. Isoperimetric inequalities relate the volume of
a manifold to its surface area. Cheeger's constant is defined to be
\begin{equation}\label{ch}
h_C = \inf_S { {\rm Vol}(S) \over {\rm min}
\{ {\rm Vol}(M_1), {\rm Vol}(M_2) \}} \, .
\end{equation}
Here $S$ runs through all compact codimension$^f$ one submanifolds which
divide $M$ into two disjoint submanifolds $M_1$, $M_2$ with common
boundary $S=\partial M_1 =\partial M_2$. A familiar example is the
two-sphere. In this case $M_1$ and $M_2$ are both hemispheres, $S=\partial
M_1$ is a great circle and we find $h_C(S^2)=1$.

Using his isoperimetric constant, Cheeger\cite{cheeg1} derived the
lower bound
\begin{equation}\label{chg}
q_1^2 \geq {h_C^2 \over 4} \, .
\end{equation}
A decade later Buser\cite{buser1} derived the upper bound
\begin{equation}\label{bs}
q_1^2 \leq 4 h_C + 10 h_C^2 \, .
\end{equation}
Cheeger's bound is valid for arbitrary closed manifolds in any
dimension. Buser provided a general bound valid in any dimension for
any closed manifold with bounded Ricci curvature. We have quoted Buser's
bound in the form relevant for 3-manifolds with constant negative curvature.
Given a 3-manifold $\Sigma$, we can in principle calculate Cheeger's
constant and subsequently use it to place bounds on $q_1$. Recasting
Cheeger's inequality in terms of wavelengths we find the maximum
wavelength is bounded from above by
\begin{equation}\label{cbnd}
\lambda_{{\rm max}} \leq {4 \pi \over \sqrt{h_C^2 -4}} \, .
\end{equation}
Similarly, Buser's inequality provides a lower bound on the maximum
wavelength:
\begin{equation}
\lambda_{{\rm max}} \geq {2\pi \over \sqrt{10h_C^2+4h_C-1}} \, .
\end{equation}
If $h_C > 2$ we would learn that $\lambda_{{\rm max}} < \infty$.
That is, there would be a long wavelength cut-off. Similarly, if
$h_C>2\sqrt{1+4\pi^2}\approx 12.7$ we would learn that 
$\lambda_{{\rm max}}< 1$ and therefore
no modes with $\lambda>1$. On the other hand, an interesting
lower bound occurs when $h_C < (\sqrt{14+40\pi^2}-2)/10 \approx 1.82$. 
In this case the manifold supports perturbations beyond the
curvature scale.

As we discuss below, the value of $h_C$ is not known for
most manifolds, but it can in principle be calculated numerically. There
are however some special examples where $h_C$ can be given a tight
upper bound. Cheeger's constant is found by simultaneously minimising 
${\rm Vol}(S)$ while maximising ${\rm Vol}(M_1)\leq {\rm Vol}(M_2)$.
The two conditions can separately be satisfied by choosing $S$ to be
totally geodesic and taking ${\rm Vol}(M_1)={\rm Vol}(M_2)={\rm
Vol}(M)/2$. One way to satisfy both of these conditions simultaneously
is to find a involution on $M$ that fixes $S$ and interchanges $M_1$
and $M_2$. Then $S$ is necessarily totally geodesic and $M_1$ and
$M_2$ are mirror images of each other. This partition
provides a local minimum for the ratio ${\rm Vol}(S) / 
{\rm min}({\rm Vol}(M_1),{\rm Vol}(M_2))$, but it might not yield the
global minimum required by the infinum in (\ref{ch}). Some interesting
examples are known where $S$ is a genus $g\geq 2$ surface and the ratio ${\rm
Vol}(S)/{\rm Vol}(M)$ is maximised, {\it ie} these examples have the
largest value of $h_C$ for manifolds that separate along a totally
geodesic boundary\cite{miya}. Written as a function of genus, the
volume of $S_g$ is $4\pi (g-1)$ and the volume of $M$ is given by
\begin{eqnarray}
&&{\rm Vol}(M_g)=g\left(-8\int_0^{\pi/4} \log|2\sin u| du
\right. \nonumber \\
&&\hspace*{0.3in} \left.
-3 \int_0^{\pi/3g} {\rm arccosh}\left({ \cos v \over 2\cos v -1}
\right)dv \right) \, .
\end{eqnarray}
For these manifolds, the bounds on $h_C$ range from $h_C\leq 1.9477$ for
$g=2$ to $h_C \leq 3.43$ in the infinite genus limit. The genus 2 case
is interesting since it tells us that there is a closed manifold with
volume $12.904..$ that supports wavelengths $\lambda >
0.94$. Furthermore, any other manifold that can be cut along a genus 2
surface will have larger volume and hence a lower isoperimetric
constant $h_C$. These manifolds will in turn support even longer
wavelength modes.

In principle it should be possible to provide a numerical estimate of
Cheeger's constant for arbitrary manifolds by trying a number
of trial partitions. The best
partitions could then be varied slightly and the search continued
until the optimal partition is found. A judicious choice for
the original trial partitions would ensure rapid convergence. The
form of Cheeger's constant (\ref{ch}) suggests that the trial
partitions should employ fairly smooth surfaces $S$ that divide
$\Sigma$ into two approximately equal sized pieces. At present no
numerical algorithm has been written, but it is hoped that the facility
will be available in later releases of the {\em SnapPea} program\cite{weeks}.

In the absence of numerical results we have to resort to analytic
estimates. An upper bound for Cheeger's constant can be derived using
geodesic balls\cite{brooks}:
\begin{equation}\label{hbnd}
h_C \leq {\partial \ln V(\chi,{\bf x}) \over \partial \chi} \, .
\end{equation}
Here $V(\chi,{\bf x})$ is the volume of a geodesic ball with radius
$\chi$ centred at ${\bf x} \in \Sigma$. The radius of the ball must be
larger than the injectivity radius$^e$ $r_{{\rm inj}}$, but
small enough so that $V(\chi,{\bf x}) < {\rm Vol}(\Sigma)/2$.
A lower bound for Cheeger's constant is quoted by Gallot\cite{gallot}:
\begin{equation}
h_C \geq { 4 \sqrt{\alpha} \over {\rm
diam}(\Sigma)(\sinh\sqrt{\alpha} + \sqrt{\alpha}) }\, ,
\end{equation}
where
\begin{equation}
\alpha \geq ({\rm diam}(\Sigma))^2\, \quad {\rm and} \quad \alpha \in
{\bf Z} \, .
\end{equation}
Notice that Eqns. (\ref{hbnd}) and (\ref{cbnd}) can be combined to
show that manifolds with diameters smaller than $0.9195$ have $q_1\geq
1$ and thus no supercurvature modes.
Applying the above bounds to Thurston's manifold (which has
$r_{{\rm inj}}=0.289$, $0.868 \leq {\rm diam}(\Sigma_{{\rm Th}})
\leq 0.88$ and ${\rm Vol}(\Sigma_{{\rm Th}})=0.9814$) we find
\begin{equation}
2.09 \leq h_C(\Sigma_{{\rm Th}}) \leq 6.42 \, ,
\end{equation}
and
\begin{equation}
1.04 \leq q_1 \leq 20.9 \, .
\end{equation}
Thus, Thurston's manifold does not support supercurvature
modes ({\it i.e.} modes with complex wavelengths), but modes
with wavelengths larger than the curvature scale are not ruled out.

Other bounds on $q_j$ exist that do not use Cheeger's constant.
Cheng\cite{cheng} provides the bound
\begin{equation}\label{ch1}
q^2_j \leq 1+ {8 (1+\pi^2) j^2 \over {\rm diam}(\Sigma)^2} \, ,
\end{equation}
and Buser\cite{buser2} provides the bound
\begin{equation}
q_j^2 \leq 1 + c\left({ j \over {\rm
Vol}(\Sigma)}\right)^{2/3} \, , \quad c > 1 \, ,
\end{equation}
but the constant $c$ is not quoted explicitly. Cheng
derived his bound by
first proving that the eigenvalues $q_j$ in a closed manifold
$\Sigma$ are always lower than the first eigenvalue of an open
geodesic ball with the same curvature and radius $\chi_0={\rm
diam}(\Sigma)/(2j)$. The bound
quoted in (\ref{ch1}) is not very sharp since Cheng considered
manifolds with arbitrary curvature. Here we derive a new, sharper bound
by specialising to three dimensional manifolds with constant negative
curvature. The first eigenvalue of an open geodesic ball of radius
$\chi_0$ is found by solving the equation $(\Delta+q^2)Q=0$
with the boundary conditions
\begin{equation}
{d Q \over d \chi}(0)=0 \, , \quad Q(\chi_0)=0 \, .
\end{equation}
The eigenfunction with lowest eigenvalue is radial ($\ell=0$, $m=0$),
\begin{equation}
Q^{q00}(\chi) = { \sin(\sqrt{q^2-1}\chi) \over \sqrt{q^2-1} \sinh\chi }
\, ,
\end{equation}
and the boundary conditions demand that
\begin{equation}
q_1^2 = 1 + {\pi^2 \over \chi_0^2}\, .
\end{equation}
From this we derive the bound on the eigenvalues of $\Sigma$:
\begin{equation}
q_j^2 \leq 1 + \left( { 2 \pi j \over {\rm diam}(\Sigma)}
\right)^2 \, .
\end{equation}
Translated into a bound on the allowed wavelengths this reads
\begin{equation}\label{db}
\lambda_j \geq {{\rm diam}(\Sigma) \over j} \, .
\end{equation}
Thus, the maximum wavelength, $\lambda_1$, is at least as large as the
diameter. The diameter is constrained to lie in the range
\begin{equation}
r_{-} < r_+ \leq {\rm diam}(\Sigma) \leq 2 r_{+} \, .
\end{equation}
Here $r_-$ is the inradius$^g$ and $r_+$ is the outradius$^e$.
The geometrical constants for a selection of {\em SnapPea}'s manifolds are
collected in Table.~II. The volume and injectivity radius are both
topological invariants while the in- and outradii depend on the choice
of basepoint for the Dirichlet domain. The diameter can be found by
forming the supremum
\begin{equation}
{\rm diam}(\Sigma) = \sup_{\bf 0} \{ r_+ \} \, ,
\end{equation}
where the supremum is take over all choices of baspoint. 

\begin{table}
\caption{Scenes from the {\em SnapPea} census.}
\begin{tabular}{ccccc}
$\Sigma$& Vol& $r_-$ & $r_+$ & $r_{{\rm inj}}$ \\
\hline 
m003(-3,1) & 0.9427 & 0.5192 & 0.7525 & 0.2923 \\
m003(-2,3) & 0.9814 & 0.5354 & 0.7485 & 0.2890 \\
s556(-1,1) & 1.0156 & 0.5276 & 0.7518 & 0.4157 \\
m006(-1,2) & 1.2637 & 0.5502 & 0.8373 & 0.2875 \\
m188(-1,1) & 1.2845 & 0.5335 & 0.9002 & 0.2402 \\
v2030(1,1) & 1.3956 & 0.5483 & 1.0361 & 0.1831 \\
m015(4,1)  & 1.4124 & 0.5584 & 0.8941 & 0.3971 \\
s718(1,1)  & 2.2726 & 0.6837 & 0.9692 & 0.1696 \\
m120(-6,1) & 3.1411 & 0.7269 & 1.2252 & 0.1570 \\
s654(-3,1) & 4.0855 & 0.7834 & 1.1918 & 0.1559 \\
v2833(2,3) & 5.0629 & 0.7967 & 1.3322 & 0.2430 \\
v3509(4,3) & 6.2392 & 0.9050 & 1.3013 & 0.1729
\end{tabular}
\end{table}

\noindent Using a more
direct numerical method we were able to compile a collection of sharp lower
bounds for the diameter. Our method ensures that the true diameter is
within $\sim 0.01$ of the lower bounds quoted in Table~III. Also
listed are upper and lower bounds on the first eigenvalue of the
Laplacian derived using the inequalities quoted in this section. 

It is interesting to note that the length of the shortest geodesic
(twice the length of the injectivity radius$^e$ $r_{{\rm inj}}$) does not
grow with the volume. Even the largest manifolds in the {\em SnapPea}
census, with volumes $\sim 6$, have geodesics as short as $0.3$ in
curvature units. This is consistent with Thurston's
assertion\cite{bill} that the expectation value for the length of
the shortest loop at an arbitrary point in a generic hyperbolic
three manifold lies in the range $0.5 \rightarrow 1$. 
This suggests that even relatively large manifolds still make for
interesting small universe models.

Having established that generic compact hyperbolic 3-manifolds support
modes with wavelengths exceeding the curvature scale, we have
partially answered the question we set out
to answer. Even neglecting the integrated Sachs-Wolfe
effect, our results show that compact hyperbolic
models are able to support the long wavelength modes required to
produce large angle anisotropy on the surface of last
scatter. A complete answer would require a knowledge of
the spectral density at long wavelengths, as a few isolated
subcurvature modes could not support significant large angle power on
the SLS. In contrast, even a single supercurvature mode $(q_1<1)$ could
greatly enhance the large angle power\cite{bell}. Preliminary results
from Bond {\it et al.}\cite{Souradeep} using the method of images
point to a reduced spectral density at long wavelengths.
Unfortunately, their method is unable to detect supercurvature
modes, so the most important part of the spectrum might be missing.

We can supplement the preceeding discussion using a theorem due to
Buser\cite{buser3} which states that there exist finite volume
compact hyperbolic 3-manifolds with an arbitrarily large number of
modes with arbitrarily long wavelength. This theorem proves that any
attempt to

\begin{table}
\caption{Diameters and eigenvalue bounds}
\begin{tabular}{cccc}
$\Sigma$& diam& $q_1$ min & $q_1$ max \\
\hline
m003(-3,1) & 0.843 & 1.08 & 7.52 \\
m003(-2,3) & 0.868 & 1.04 & 7.31 \\
s556(-1,1) & 0.833 & 1.09 & 7.61 \\
m006(-1,2) & 1.017 & 0.82 & 6.26 \\
m188(-1,1) & 0.995 & 0.84 & 6.40 \\
v2030(1,1) & 1.082 & 0.77 & 5.90 \\
m015(4,1)  & 0.923 & 0.98 & 6.88 \\
s718(1,1)  & 1.439 & 0.53 & 4.48 \\
m120(-6,1) & 1.694 & 0.45 & 3.84 \\
s654(-3,1) & 1.946 & 0.36 & 3.38 \\
v2833(2,3) & 1.701 & 0.45 & 3.83 \\
v3509(4,3) & 1.802 & 0.39 & 3.63
\end{tabular}
\end{table}

\noindent exclude {\em all} compact hyperbolic models on the basis of a lack
of long wavelength power is doomed to failure. Admittedly, the manifolds
considered by Buser have large diameters, but they also have small
injectivity radii so they describe models that are multi-connected on
scales smaller than the curvature scale. The fundamental cells for
these manifolds are highly anisotropic, which may bring them into
conflict with observations, but this is not certain since the face
identifications tend to mix all three spatial directions and thus
apparent isotropy can be restored.

\section{Horned Topologies}

In this section we digress to consider a particular class of models
that can be partially constrained by COBE data.
In Ref.\cite{janna}, Levin {\it et al.} describe the microwave sky in
a universe with the topology of a hyperbolic toroidal 
horn. The topology they consider is the three dimensional analogue of
the two dimensional pseudosphere. The pseudosphere, refered to as a cusp by
mathematicians, is topologically equivalent to
$S^1 \times [0,\infty)$, where $S^1$ is a circle.
Figure 4 shows a portion of the pseudosphere embedded in three
dimensional space. The pseudosphere is 
described in the upper half plane representation of $H^2$ by
\begin{equation}
d\sigma^2 = { dx^2 + dz^2 \over z^2},
\end{equation}
with the identifications $x=x+n L_x$ with $n \in \bf{Z}$.
\
\begin{figure}[h]
\vspace{58mm}
\includegraphics{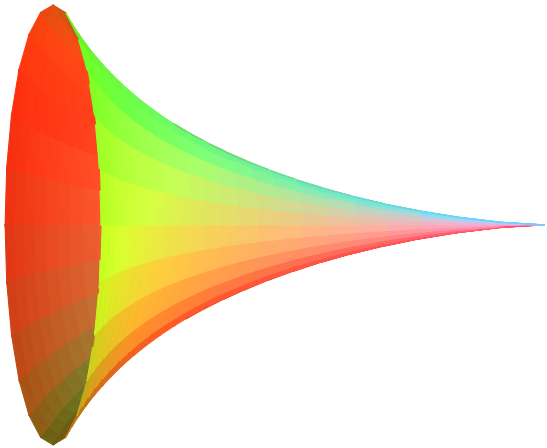}
\vspace{8mm}
\caption{A portion of the 2-d pseudosphere embedded in 3-d space.}
\end{figure} 
\noindent Cusps in $d$ dimensions
are analogously defined to be of the form $E^{d-1} \times [0,\infty)$
where $E^{d-1}$ is a flat topology in $(d-1)$ dimensions. It should be
emphasised that the line $z={\rm const.}$ connecting $x$ and $x+L$ is
not a geodesic. Geodesics in the upper half plane model appear as
half-circles of the form $x^2+z^2=a^2$, perpendicular to the boundary plane.

The hyperbolic horn studied in Ref.\cite{janna} is of the form
$T^2 \times [0,\infty)$ where $T^2$ is the two-torus. In the upper
half plane model of $H^3$,
\begin{equation}
d\sigma^2 = { dx^2 + dy^2+ dz^2 \over z^2},
\end{equation}
the horn is defined by making the identifications $x=x+nL_x$ and
$y=y+mL_y$. Since translations commute, the horn's fundamental group
is abelian and geodesics on the horn are non-chaotic. This means that
the horn's eigenmodes can be written down explicitly.

The calculational simplicity of the horn model is offset by some
unappealing physical characteristics. Not only is the horn
noncompact and infinite in volume, but it also suffers from
severe global anisotropy.
The anisotropy can be seen by moving to
spherical coordinates centred at $(x,y,z)=(0,0,1)$. Our
ghost images then appear at the points
\begin{eqnarray}
&&\chi={\rm arcsinh}\left({\rm arccosh}\left(1+\frac{1}{2}(n^2
L_x^2+m^2L_y^2) \right)\right) \, , \nonumber \\
&&\theta=\pm {\rm arccos}\left(\left[1+{4 \over (n^2
L_x^2+m^2L_y^2)}\right]^{-1/2}\right) \, , \nonumber \\
&&\phi={\rm arcsin}\left({ m L_y \over \sqrt{n^2 L_x^2+m^2L_y^2}}
\right) \, .
\end{eqnarray}
The ghost images are evenly distributed in the $\phi$ direction, but
distant images pile up along the axis of the horn ($\theta=\pi/2$).

Because the horn's fundamental group is abelian, there will be a long
wavelength cut-off in directions orthogonal to the axis of the
horn. In this respect the hyperbolic horn is similar to the flat
topology $\Sigma=R\times T^2$. The difference is that the torus
cross-sections of the horn do not have fixed area. As we
move away from the origin, the torus area decreases like
\begin{equation}
A(T^2)=\frac{1}{2} L_x L_y \exp(-2\sinh\chi) \, .
\end{equation}
This means that the wavelength cut-off gets shorter and shorter as we
move toward the cusp. Moreover, the
decrease is {\em doubly exponential} with increasing proper distance. This
has the effect of suppressing all temperature fluctuations in the
direction of the horn, leading to a ``flat-spot''\cite{janna} in the
microwave sky.

Considering that there are an infinite number of hyperbolic three
manifolds to choose from, it might seem strange to focus on one
particular example. However, it turns out that many manifolds have
horn-like regions. To see why one needs to understand something about
how hyperbolic three manifolds are constructed.  According to
J{\o}rgensen's theorem\cite{jorg}, all finite volume hyperbolic three
manifolds can be obtained by Dehn surgery on a finite number of link
complements in $S^3$. A link compliment is constructed by drilling out a solid
tubular knot or link from spherical space. The complement of this link
({\it ie.} the space outside the link) will almost always be
topologically equivalent to a hyperbolic three manifold with one or
more cusps. If one happened to live deep inside a cusp, the universe
would look exactly like a toroidal horn. 

While the finite volume of
the cusped manifolds makes them more appealing than the
basic hyperbolic horn, they are still non-compact. In order to arrive
at compact models we need to perform Dehn surgery on the link. The
surgery involves cutting out a portion of the link and replacing it
with a solid torus that is first twisted around the link in some
non-trivial way. Without going into details\footnote{See Thurston's
book\cite{billbook} or the appendix of Carlip's article\cite{steve}
for a description of how to perform Dehn surgery}, it is
sufficient to note
that the twisting can be parametrised by two integers $(p,q)$. In the
limit $p,q \rightarrow \infty$ with $p,q$ relatively prime, the
original cusped manifold is recovered. For small $p,q$ the cusp can be
completely removed. For large values of $p,q$ the end of the cusp is
rounded off leaving a ``horned manifold''. If one happened to live deep
inside one of the horns, the universe would look similar to how it
does in the infinite toroidal horn. The exact correspondence is broken
since Dehn surgery makes the fundamental group non-abelian. This means
that the geodesics will be chaotic and the eigenmodes
complicated. Nonetheless, for high order Dehn fillings the chaos should
be mild and it seems reasonable to expect a flat spot in the CMB if
one lived in a horned region of the manifold.

The preceeding considerations have shown that the results of
Ref.~\cite{janna} apply in certain regions of
a large class of three manifolds. If we happened to live in one of
these horned
regions, we would see a severe suppression of CMB fluctuations along
the horn. Levin {\it et al.} found that this effect was not masked by
the integrated Sachs-Wolfe effect, so the COBE satellite would have
detected flat spots in the CMB. However, the absence of flat spots is
not a very strong constraint on us living in a cusped manifold. This
is because cusps only account for a very small portion of a
cusped manifold's volume. Therefore, it is very unlikely that we would
be living in or near a cusp.
If we make what topologist refer to as a ``thick-thin''
decomposition\cite{billbook}, we find that most of a manifold's volume
is in the ``thick'' part and very little is in ``thin'' regions such as
cusps. The chance that we live deep inside a cusp is even smaller
since the volume of a cusp decreases as $\exp(-2 e^\chi)$, where
$\chi$ is the proper distance down the cusp. We are far more likely to
live in a thick portion of a manifold where the breaking of
global isotropy is much less noticeable. The analysis of Levin {\it et
al.} is not valid for observers that live in the thick
portion of a manifold. The fundamental group of a cusped manifold
is non-abelian, but contains a normal abelain subgroup of finite index,
corresponding to isometries of the cusp. Inside a cusp the fundamental
group is dominantly abelain and the horn analysis holds, but in the thick
part of the manifold the isometries are dominantly non-abelian and the
horn analysis does not hold. It would be interesting to extend the
horn analysis to cusped manifolds with finite volume and to closed
manifolds with horn-like regions.

In summary, it would be
surprising if we did live in a horned region, and the results of
Levin {\it et al.} confirm that we do not!

\section{Power Spectra, Wavenumbers and Multipoles}

\subsection{Generating the primordial power spectrum.}

The temperature fluctuations measured by COBE-DMR are thought to arise
from the amplification of quantum fluctuations during an inflationary
phase, or alternatively, from a network of topological defects.
We will not consider the latter possibility as there appears to be
a topological obstruction to the formation of topological defects
in a small universe\cite{uzan}. In the inflationary context, some fine
tuning is required to avoid blowing the curvature scale outside the
surface of last scatter. There are currently two scenarios for
arriving at a negatively curved
universe from inflation. The first is one-bubble inflation\cite{gott},
the second is compact inflation\cite{css1}. Detailed calculations of
the power spectrum have been performed for the one-bubble model, while
little is known about the spectrum for compact inflation. Here we
discuss how the one-bubble scenario relates to multiconnected models,
and offer some speculations about the form of density perturbations
produced by compact inflation.

\subsubsection{One-bubble inflation}

Since there is an explicit and well understood quantum tunnelling
process underlying the one-bubble inflation scenario, it is possible to
make definite predictions about the form of the primordial power
spectrum\cite{gott}. The universe begins in an inflationary epoch
driven by an inflaton field
in a false vacuum state. During this epoch, any inhomogeneities are
inflated away. Subsequently, a single bubble is nucleated, inside of which 
the inflaton field rolls toward its true minimum. Taking the inflaton
to be described by a single real scalar field $\phi$ (several variants
of this basic picture have been considered), surfaces of constant
$\phi$ inside the bubble have constant negative
curvature. Mathematically this process is described by an $O(4)$
symmetric Euclidean instanton -- Euclidean de Sitter space with one
special point. The bubble nucleation selects a
prefered point in de Sitter space, breaking the full $O(5)$ symmetry down
to $O(4)$. As shown in Fig.~5, the Euclidean instanton is matched
onto its Lorentzian counterpart across a totally geodesic spatial
hypersurface, $S^3=\partial S^4$. The matching surface is a Cauchy
surface for the subsequent Lorentzian evolution. Owing to the
$O(4)$ symmetry of the instanton, the bubble interior has the
$SO(3,1)$ symmetry of hyperbolic space.

\
\begin{figure}[h]
\vspace{40mm}

\includegraphics{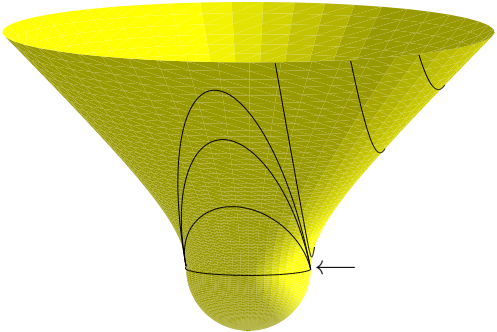}

\vspace{2mm}
\caption{The tunnelling configuration. The lines indicate constant time
hypersurfaces. The arrow indicates the point on the $S^3$ Cauchy
surface where the bubble nucleates. The region to the upper right of this
point is the interior of the bubble, the region to the upper left remains in
the false vacuum.}
\end{figure} 

\vspace*{-2mm}
\begin{picture}(0,0)
\put(137,100){$\longleftarrow$}
\end{picture}

The question we need to ask is: can the one-bubble scenario be
generalised to produce a hyperbolic universe with non-trivial spatial
topology? The answer is no, unless one is willing to live with closed
timelike curves. The only way non-trivial topology can enter into the
picture is through the spatial topology of $S^3$,
since the Lorentzian evolution is fully determined by specifying
Cauchy data for the metric and matter fields on the initial $S^3$
Cauchy surface. Put differently, the
solution is specified globally on the $S^3$ hypersurface but only
locally in the direction normal to this surface. Let us start by
considering the simplest non-trivial topology for the matching
surface -- the real projective space $RP^3\cong S^3/Z_2$. Taking the
geometry shown in Fig.~5 as the universal cover, we find there is a
clone of the bubble on the opposite side of de Sitter
space. Since antipodal points in de Sitter space lie outside each
other's light cone, the bubble and its clone never intersect. Inside
the bubbles we have two copies of the same simply connected hyperbolic
universe. While this sounds reasonable, we encounter a problem when
trying to define quantum fields in this background as the
instanton has topology $RP^4$ and is thus non-orientable. It is
impossible to separate modes into positive and negative frequency
components in such spacetimes. As we move on to consider more
complicated topologies the situation gets worse.
Once the number of clones exceeds two, the bubbles start to
collide (self-intersect). Moreover, according to an observer
inside the bubble, the spatial
identifications on the spherical slicings become spatio-temporal
identifications in hyperbolic space. These universes have 
closed time loops and there is no known prescription for defining a
sensible quantum theory in such spacetimes.

So while the most complete calculations of the primordial
power spectrum have been done in the context of one-bubble inflation, they
can not be generalised to models with compact hyperbolic
sections. Indeed, if we do find evidence for non-trivial (purely) spatial
topology, we would know that the one-bubble model is ruled out.

\subsubsection{Compact inflation}

In the compact inflation scenario the universe is taken to have
compact hyperbolic spatial sections. The chaotic mixing that occurs in
compact hyperbolic space is understood to have erased any initial
density perturbations before vacuum domination is reached\cite{css1}.
This prepares the ground for a necessarily short burst of inflation.
It is the chaotic mixing that solves the horizon problem,
and not the short period of inflation.

In order to calculate the quantum fluctuations in compact inflation we
need to know the eigenmodes and the initial vacuum state, {\it ie.}
how the modes are populated. Once these are known we can evolve
the quantum fluctuations to find the density perturbations at the end
of inflation. At present we know neither the eigenmodes nor the
correct vacuum state to choose. The situation is only slightly better
for non-compact open inflation models, for while the eigenmodes
are know, the choice of initial vacuum is not. Here we are referring to
models other than the one-bubble model (see {\it eg.}
Ref.~\cite{lyth}). Another difficulty with open inflation stems from the
short period of inflation being insufficient to solve the horizon problem.
This problem is greatly exacerbated by the open universe
Grishchuk-Zel'dovich effect\cite{bllw}, which demands that the
pre-inflationary universe be smooth on scales one thousand times
larger than the curvature scale. In a compact hyperbolic universe this
is not a problem since the entire universe is typically no larger than
the curvature scale. In a small universe there can be no ``monsters''
lurking over the horizon, for we already see all there is to see.

The generation of curvature perturbations in an open universe closely
parallels that in a flat universe: Starting in some initial state,
the perturbations evolve until they cross outside the Hubble horizon,
whence they are frozen in. After inflation, the Hubble horizon expands to
encompass perturbations of increasing wavelength. Once back inside the
Hubble radius the perturbations can undergo further evolution.
In an open universe the large scale perturbations are amplified by the ISW
effect. Fluctuations with wavelengths much smaller than the curvature scale
are insensitive to the curvature and evolve just as they do in the
flat models. For these modes it seems reasonable to use the usual
conformal vacuum initial conditions\cite{lyth}.
We anticipate the same should be true in compact
hyperbolic models for fluctuations frozen in long after the topology
scale exited the Hubble horizon. The situation is far more complicated
near the curvature scale, as it is here that the geometry and
topology of the background manifold become most important. On large
angular scales there will be a delicate interplay between 6 main
effects: (1) Amplification by the ISW effect; (2) Suppression by
gravitational focusing; (3) Curvature distortions of the conformal
vacuum; (4) Inflationary transients;  (5) Finite size distortions
of the conformal vacuum; (6) Reduced spectral density at long
wavelengths. The first four of these effects are present in all open
inflation models while the last two are unique to compact inflation.
Taking into account effects (1) and (2) while neglecting effects (3)
and (4) leads to an angular power spectrum with positive
slope\cite{silk}. Of the remaining effects, it is likely that
inflationary transients will boost the power on large scales while any
reduction in the spectral density will tend to reduce power on large
angular scales. The hardest questions to answer concern distortions to
the conformal vacuum, as these can only be answered in the
context of quantum cosmology by repeating the sort of calculations
performed by Halliwell and Hawking for closed models\cite{hh}.
Until this is done, and until more is known about the eigenmode
spectra, no firm predictions can be made about the power spectrum on
large angular scales.

\subsection{Wavenumbers and Multipoles}

In section II we described how the ISW effect alters the flat space
relationship between wavenumber $k$ and multipole number $\ell$. In flat
space, the dominant contribution to the $\ell^{\rm th}$ multipole
comes from fluctuations with wavenumber $k_{sls} \sim \ell +1$. This
simple relationship between $k$ and $\ell$ translates into
a simple relationship between the power spectrum of density perturbations,
$P(k)=|\delta_k|^2$, and the rms temperature fluctuations in each multipole,
$P(l)=(\ell(\ell+1)C_\ell)^{1/2}$, where $C_\ell=<\! |a_{\ell m}|^2\!
>$. The standard
example is a flat universe with power-law spectrum $P(k)\propto
k^n$, where it is found that
\begin{equation}
C_\ell \propto { \Gamma[3-n] \over \Gamma[(4-n)/2]^2 } 
{\Gamma[(2\ell+n-1)/2] \over \Gamma[(2\ell+5-n)/2]} \, .
\end{equation}
Already we have seen that the ISW effect breaks this correspondence on
large angular scales in an open universe. In compact
hyperbolic space there is an additional effect that tends to spread
power from different scales into each multipole. Physically,
this effect is related to quantum chaos and the mixing properties of
compact hyperbolic spaces. Mathematically, the effect arises because a
multipole expansion attempts to use a basis of smooth analytic functions to
represent the non-analytic wavefunctions. According to
Berry\cite{berries}, if $\Psi_k({\bf x})$ is an eigenmode of $H^3/\Gamma$,
then the coefficients $a_{\ell m}$ in the expansion
\begin{equation}\label{expan}
\Psi_k({\bf x})=\sum_{\ell}\sum_{m=-\ell}^{m=\ell} \delta_k \,
a_{\ell m} X^{\ell}_{k}(\chi) Y^{\ell}_{m}(\theta, \phi) \, ,
\end{equation}
are essentially random variables with amplitude virtually independent
of $\ell$. At small wavenumbers and large angular scales, there is 
almost no correlation between $k$ and $\ell$ in compact hyperbolic space.
The convergence of the sum (\ref{expan}) is inherently slow since 
it comes not from a decrease in
the expansion coefficients, but from the decay of $X^{\ell}_{k}(\chi)$
across the fundamental domain.

Some concrete results are known in two dimensions that nicely
illustrate this effect. Consider a genus 2 (hyperbolic) surface with
eigenmodes $\Psi_k({\bf x})$. The analogue of (\ref{expan}) is then
\begin{equation}
\Psi_k({\bf x})=\sum_{m=-\infty}^{m=\infty} \delta_k \,
a_{m} Y^{m}_{k}(\chi)\, e^{i m \phi} \, .
\end{equation}
Here it is known that
\begin{equation}
c_1 |m|^{-1/2} \leq |a_m| < c_2 |m|^{1/2} \, ,
\end{equation}
where $c_1$ and $c_2$ are constants. It is thought that
the lower bound $|a_m| \sim |m|^{-1/2}$ is
a good estimate of the true behaviour\cite{balazs}. This leads to
an essentially flat angular power spectrum\footnote{$P(m)$ is
the 2-dimensional analogue the of the usual angular power spectrum,
$P(\ell)$, in 3-dimensions.} $P(m)=(2|m|)^{1/2}|a_m|
= {\rm const.}$, regardless of the perturbation spectrum $\delta_k$.

The above results suggest a novel way of arriving at a nearly flat
Harrison-Zeldovich spectrum, regardless of the underlying physical
process that produces the fluctuations. We expect that this
redistribution of power will be most efficient on large angular
scales and least efficient on small angular scales. Our reasoning is
that long wavelength modes are the hardest to approximate by
analytic functions since they are most affected by the complicated
periodic boundary conditions imposed by the topology. Conversely, the
short wavelength modes are less sensitive to global effects, and
therefore well approximated by the corresponding eigenmodes of
infinite hyperbolic space.

\section{conclusions}

A hyperbolic drum produces a rich and complex sound. A compact
hyperbolic universe is likewise infinitely more complex than its
spherical or euclidean counterparts. The simple methods used to
constrain flat models do not work when space is negatively curved.
The eigenmodes in a compact hyperbolic space can only be calculated
using sophisticated methods developed to treat quantum chaos.
Moreover, hyperbolic models do not suffer the simple long wavelength
cut-off used to exclude toroidal models.

In addition to the issue of what fluctuations are supported on the last
scattering surface, there is also the issue of what exactly it was that
COBE measured.
In a compact hyperbolic universe the curvature radius provides a
natural length scale, $R_0=H_0^{-1}/\sqrt{1-\Omega_0}$.
The curvature radius sets the length scale where we might hope to find
the first evidence that we live in a multiply connected universe.
The curvature radius also sets the angular scale beyond which
fluctuations in the cosmic microwave background radiation no
longer originate from the last scattering surface. This confluence of
physical scales is very unfortunate for COBE since it means that the
ISW effect takes over just when things get interesting. Fortunately
the next generation of CMB satellites will be able to probe much
smaller angular scales, so the ISW effect will not obscure their view
of the large scale topology of the universe. 

The search for multi-connectedness in our universe is not over. It
has barely begun.

\section{Acknowledgements}

We are indebted to Jeff Weeks for his extensive expert advice on topology and
the workings of {\em SnapPea}. Our knowledge of cusped manifolds was
greatly enhanced by discussions with Bill Thurston and Jeff Weeks.
We thank the topologists
Bob Brooks, Pat Callahan, Ruth Kellerhals and Alan Reid for sharing
their knowledge on the eigenvalue spectra. We have also enjoyed
informative discussions with Dick Bond, Andrew
Chamblin, Fay Dowker, Gary Gibbons and Janna Levin.
N. Cornish was supported by PPARC grant GR/L21488.
D. Spergel was supported by the NASA grant for the Microwave Anisotropy Probe.
G. Starkman was supported by a NSF career grant.

\section*{appendix}

The following glossary of terms describe the basic mathematical
quantities used in our discussion of topology. Our definitions are
designed to be more pictorial than the usual formal definitions
found in the mathematical literature.\\
$[a]$ {\bf Betti numbers}: The zeroth betti number, $b_0$, counts the number
of disconnected regions in a manifold. The first betti number, $b_1$,
counts the number of incontractible loops. The second betti number,
$b_2$, counts the number of incontractible surfaces. Higher betti
numbers are similarly defined. The first betti number is equal to the
rank of the free abelian part of the first homology group
$H_1(\Sigma)$. In other words, the first betti number is equal to
the number of generators of $H_1(\Sigma)$ that are not subject to any
relations save those that make the group abelian. Poincare duality
relates the various betti numbers so that in $d$-dimensions
$b_i=b_{(d-i)}$.\\
$[b]$ {\bf Injectivity radius function}: The injectivity radius of a
point $p\in M$, $r_{{\rm inj}}(p)$, is the radius of the largest
coordinate chart that can be centred at $p$. Since a coordinate chart
breaks down when any geodesic refocuses, the injectivity radius of a
point is half the length of the shortest geodesic connecting $p$ to
itself.\\
$[c]$ {\bf Injectivity radius}: The injectivity radius of a manifold,
$r_{{\rm inj}}(M)$, is the smallest injectivity radius of any point
in the manifold, {\it ie.}
\begin{equation}
r_{{\rm inj}}(M) = \inf_p r_{{\rm inj}}(p\in M)\, . 
\end{equation}
Thus, $r_{{\rm inj}}(M) = l_{{\rm min}}/2$, where
$l_{{\rm min}}$ is the length of the shortest geodesic in $(M,g)$.\\
$[d]$ {\bf Diameter}: The diameter, ${\rm diam}(M)$, of a manifold
$(M,g)$ is the greatest distance between any two points on the manifold.\\
$[e]$ {\bf Outradius}: The outradius, $r_{+}$, of a compact hyperbolic
manifold fixes the size of the smallest hyperbolic ball
that can be used to enclose the fundamental cell.\\
$[f]$ {\bf Codimension}: The codimension is a complimentary dimension. A
$n$-dimensional hypersurface living in a $d$-dimensional space has
codimension $d-n$. \\
$[g]$ {\bf Inradius}: The inradius, $r_{-}$, is the radius of the largest
simply connected geodesic ball in $(M,g)$. In other words, $r_{-}$ is
the largest distance any point in the manifold can be from its closest
image. For a compact hyperbolic
manifold, the inradius fixes the size of the largest hyperbolic ball
that can be placed inside the fundamental cell.

\end{document}